\documentclass[conference]{IEEEtran}
\IEEEoverridecommandlockouts
\usepackage{cite}
\usepackage{amsmath,amssymb,amsfonts}
\usepackage{algorithmic}
\usepackage{graphicx}
\usepackage{textcomp}
\usepackage{xcolor}
\usepackage{balance}
\usepackage{enumitem}
\def\BibTeX{{\rm B\kern-.05em{\sc i\kern-.025em b}\kern-.08em
    T\kern-.1667em\lower.7ex\hbox{E}\kern-.125emX}}

\newcommand*\rot{\rotatebox{90}}

\begin{document}

\title{Tailoring Education with GenAI: A New Horizon in Lesson Planning
\thanks{This work has been funded by the Horizon Europe Road-STEAMER project (Grant number: 101058405). More information can be found at https://www.road-steamer.eu}}

\author{\IEEEauthorblockN{Kostas Karpouzis}
\IEEEauthorblockA{\textit{Dept. of Communication, Media and Culture}\\
\textit{Panteion University of Social and Political Sciences}\\
Athens, Greece \\
kkarpou@panteion.gr} \and
\IEEEauthorblockN{Dimitris Pantazatos}
\IEEEauthorblockA{\textit{National Technical University of Athens}\\
\textit{Network Management \& Optimal Design Laboratory} \\
Greece \\
dpantazatos@netmode.ntua.gr} \and
\IEEEauthorblockN{Joanna Taouki}
\IEEEauthorblockA{\textit{Instructional Design Team,} \\
\textit{100mentors eLearning Providers}\\ 
Athens, Greece \\
joanna.taouki@100mentors.com} \and
\IEEEauthorblockN{Kalliopi Meli}
\IEEEauthorblockA{\textit{Dept of Educational Sciences \& Early Childhood Education} \\
\textit{University of Patras}\\
Greece \\
kmeli@upatras.gr} 
}

\maketitle

\begin{abstract}
The advent of Generative AI (GenAI) in education presents a transformative approach to traditional teaching methodologies, which often overlook the diverse needs of individual students. This study introduces a GenAI tool, based on advanced natural language processing, designed as a digital assistant for educators, enabling the creation of customized lesson plans. The tool utilizes an innovative feature termed 'interactive mega-prompt,' a comprehensive query system that allows educators to input detailed classroom specifics such as student demographics, learning objectives, and preferred teaching styles. This input is then processed by the GenAI to generate tailored lesson plans.
To evaluate the tool's effectiveness, a comprehensive methodology incorporating both quantitative (i.e., \% of time savings) and qualitative (i.e., user satisfaction) criteria was implemented, spanning various subjects and educational levels, with continuous feedback collected from educators through a structured evaluation form. 

Preliminary results show that educators find the GenAI-generated lesson plans effective, significantly reducing lesson planning time and enhancing the learning experience by accommodating diverse student needs. This AI-driven approach signifies a paradigm shift in education, suggesting its potential applicability in broader educational contexts, including special education needs (SEN), where individualized attention and specific learning aids are paramount. \end{abstract}


\section{Introduction}Traditional educational systems, with their uniform teaching methodologies, often fail to address the diverse needs of individual students. This uniformity leads to a one-size-fits-all approach, which neglects the students’ unique learning styles, cultural backgrounds, and personal interests \cite{b1}. The advent of Generative AI (GenAI) in education presents an opportunity to overcome these limitations by enabling the customization of learning materials to suit individual or group student needs better, fostering a more engaging and effective learning environment, and aligning with contemporary educational goals \cite{b2}, \cite{b3}.
The scope of this work encompasses exploring the efficacy and adaptability of Generative AI in personalizing education. It delves into how GenAI can be tailored to meet diverse learning styles and needs, enhancing the educational experience. Furthermore, the present study aims to critically assess the role of GenAI in lesson planning through quantitative and qualitative analyses, considering its impact on various educational levels and subjects. This exploration underlines the potential of GenAI as a transformative tool in educational practices and future research.
The paper is structured as follows: Following the introduction (Section I), Section II discusses personalization in education, focusing on the adaptability of GenAI for individual learning needs. Section III covers related work in the field, setting the theoretical foundation. Section IV elaborates on the methodology, emphasizing the development of interactive prompts with GenAI. Section V includes the results based on the evaluation of the learning scenarios, and the analysis of the effectiveness of GenAI in educational contexts. Finally, Section VI, concludes with a discussion and our conclusion on the impact and future potential of GenAI in education.

\section{Personalization in education}
The advent of Generative AI (GenAI) in education, particularly in the sphere of personalized learning, represents a significant leap forward. These technologies, such as OpenAI's GPT models, are not just automating aspects of the educational process; they are redefining it. By offering a more tailored educational experience that aligns with the unique needs, learning styles, and pace of individual students or groups of students, GenAI promises a more engaging and effective learning experience.

A major benefit of GenAI in personalized learning lies in its ability to create adaptive learning paths. Unlike traditional one-size-fits-all approaches, GenAI can analyze individual student performance and preferences, dynamically adjusting its content and resources to suit each learner's unique requirements. For example, a GenAI system might detect students struggling with algebra and automatically introduce more foundational math concepts before progressing, or it might present advanced science topics in an interactive format to students showing a keen interest and aptitude in the subject. This adaptability caters to students' diverse needs and promotes a more effective and engaging learning experience. This concept is eloquently explored in \cite{b10}, which underscores the adoption and potential of AI in crafting personalized learning experiences.

Another significant advantage of GenAI-driven personalized learning is the heightened level of student engagement it fosters \cite{b2}, \cite{b7}. GenAI tools can ignite a deeper interest and motivation in learners by aligning educational content with students' interests and proficiency levels. Furthermore, GenAI enhances accessibility in education, a crucial aspect in today's diverse learning environments. By providing a variety of learning materials and adapting to different learning preferences and needs, GenAI tools make education more inclusive and accessible to a broader spectrum of students. This aspect is particularly emphasized in \cite{b8} and \cite{b16}, which highlight the role of AI in making learning resources more accessible to diverse learner populations.

\subsection{Adaptive lesson planning}
To enable students to experience the potential of GenAI in their everyday learning, it is paramount that educators design and implement their lessons to accommodate their student’s unique cognitive needs. When it comes to lesson planning, the application of GenAI is reshaping the landscape. These systems offer the potential not only to automate but also to enhance the lesson-planning process. By analyzing comprehensive curricula and diverse student data sets, GenAI can develop dynamic, customized lesson plans that can be continuously refined based on ongoing student feedback and performance data \ref{b15}. This approach promises a more responsive and effective educational experience \cite{b9}, illustrating the transformative impact of AI on the efficiency and effectiveness of curriculum development.

However, this technological advancement is not without its challenges. One of the primary concerns is the potential over-reliance on technology in lesson planning. There's a risk that educators might become too dependent on AI tools, potentially diminishing their role in the creative and professional aspects of lesson planning. \cite{b11} delves into this issue, emphasizing the need for educators to maintain a significant role in the lesson planning process, ensuring that the use of AI complements rather than replaces their expertise and creativity.

\subsection{Personalized evaluation}
To evaluate the impact of each lesson plan and its implementation, regardless if it is GenAI-augmented or not, it is crucial to assess the consequent learning progression of participating students. In the area of evaluation and assessment, GenAI is bringing about transformative changes. GenAI systems can provide immediate and tailored feedback such as pinpointing specific areas of misunderstanding in a student's answer, suggesting tailored resources for improvement, or offering real-time corrections during interactive learning sessions. Minn \cite{b12} discusses how instant feedback provided by AI can accelerate the learning process and significantly enhance student understanding and performance.

Despite these advantages, the application of GenAI in student evaluation raises critical issues, notably regarding data privacy and the potential for bias. The use of detailed student data by AI systems for assessment purposes brings to the forefront significant concerns about privacy and data security, as explored in \cite{b13}. Moreover, the possibility of AI algorithms perpetuating existing biases, leading to unfair and skewed assessments, is a critical concern that must be addressed. Gaskins in \cite{b14} provides an in-depth examination of this issue, highlighting the need for careful and ethical implementation of AI in educational assessments.

\section{Prompt Engineering}
The emergence of artificial intelligence (AI) prompt engineering as a new digital competence has garnered significant attention in various domains, including design research, scientific authorship, and education. According to P. Korzynski et al. \cite{m2}, AI prompting can be considered a new digital competence. To further strengthen their opinion, they provide a theoretical framework for optimal approaches in AI prompt engineering and introduce the AI PROMPT framework, guiding text-to-text prompt engineering. The AI PROMPT framework is based on seven key aspects, which are:
\begin{itemize}
    \item Articulate the Instruction
    \item Indicate the Prompt Elements
    \item Provide Ending Cues and Context
    \item Refine Instructions to Avoid Ambiguity
    \item Offer Feedback and Examples
    \item Manage Interaction
    \item Track Token Length and Task Complexity
\end{itemize}
This framework was utilized on ChatGPT and Google Bard prompts. 

Few-shot prompting is a method for providing a few examples of a specific task. In this approach, the GenAI tools takes this text as input and generates the answer or ranks different options \cite{m3} in contrast to the zero-shot prompting, which involves text generation without training on the specific task \cite{m4}. This method can be more scalable and sometimes requires relatively few input data \cite{m5}. Although this method can be considered a common way to help genAI models understand and cope with a task, even this method can be further improved. One exciting work in this direction is that of Brian Lester et al. \cite{m6}, which involves learning “soft prompts'' that condition the model to perform specific tasks without modifying the core model parameters. This method seems to outperform the few-shot learning capability of ChatGPT-3 and is more efficient, but it is not as simple as the few-shot method. 

A challenge related to prompt creation is when non-AI experts attempt to design prompts for these models. As J.D. Zamfirescu-Pereira et al. \cite{m7} highlighted, non-experts often struggle with understanding and creating effective prompts for large language models (LLMs). They tend to overgeneralize from limited experiences and work to evaluate prompt effectiveness, revealing a prominent gap in prompt design literacy among non-experts. This underscores the necessity for more intuitive tools and educational resources to aid in the prompt design process, especially as these models become more prevalent and accessible to a broader range of users.

The aforementioned approach is further addressed in the study of  M. Sharples \cite{m8}. Sharples suggests a more conversational and interactive form of engagement between students and AI, transcending traditional prompt-response methods. This approach could potentially bridge the gap in prompt design literacy among non-experts by enabling AI to participate more dynamically in educational settings, fulfilling roles like co-designers or storytellers. It aligns with the need for AI systems that are both technologically advanced, pedagogically supportive, and ethically conscious, enhancing the learning experience and making AI interactions more accessible and effective for all users.

In the following section, we present the methodology behind the creation of the Learning Scenario assistant.

\section{Methodology}
Building upon the theoretical foundations and prompt engineering strategies discussed in the previous section, we hereby introduce our cutting-edge methodology for constructing interactive prompts with GenAI tools. Moving beyond standard AI practices, our approach engages GenAI in a user-centric dialogue thoughtfully designed to exchange prompts and responses (follow-up prompts). This exchange is intended to navigate GenAI through complex workflows by replicating the conversational dynamics of human engagement. Our method establishes a two-way communication channel, allowing GenAI to interact with users, building upon their inputs with progressive dialogue to refine solutions collaboratively.

This sophisticated approach transforms GenAI from a mere tool into an interactive partner in task management and execution. By providing clear, context-rich, and purpose-driven prompts, we aim to enable users to effectively direct the GenAI tool to deliver more relevant and high-quality results, thus enhancing their overall workflow efficiency and productivity. The subsequent paragraphs will delve into the specific elements of the interactive prompt methodology, illustrating its application in creating a learning scenario and underscoring the structured, dialogic approach that can be adapted to meet the unique needs of educational environments.

\subsection{Sections of an Interactive Prompt}
These are the critical elements of the proposed methodology for creating an Interactive Prompt:
\begin{itemize}
    \item Positioning Prompt(s) (“Gradual Build Up”): These initial prompts aim to orient the Large Language Model (LLM) within a specific domain, analogous to setting the scene in a conversation between humans (e.g., What are some good practices for creating a learning scenario for students of various learning needs?). This step is followed to ensure that the LLM is accurately positioned in the intended area of focus. This part of the prompting process can be considered as a modified few-shot prompting technique, as the user can evaluate the LLM’s answer/output of the positioning prompt and provide additional, targeted information to fine-tune its understanding.
    \item Interactive Prompt: The Interactive prompt instructs the LLM to query the user and gather essential information for performing a specific task— in this case, creating a learning scenario or lesson plan for students. It systematically inquires about key components like the target audience, project context, and the specific task at hand, alongside objectives and format preferences. This process mirrors a conversation between humans, where one party seeks detailed input to understand and fulfill the requirements of the other, ensuring that the AI can create customized content – here, a tailored educational experience.
    \item Follow-up Prompt: This prompt is crafted by the user aiming to follow up on the Interactive Prompt's output, in order to delve deeper into the specifics provided by the LLM, asking for clarification, refinement, or additional instructions on how to improve task execution. This step is akin to a dynamic human dialogue, where each query is aimed at sharpening understanding and moving towards the desired outcome progressively. 
    \item Practical Application and Evaluation: The final step involves the practical application and evaluation of the Prompt Thread with end-users, incorporating quantitative analysis and quality assessment.
\end{itemize}
The next section transitions from the theoretical framework to a concrete case study, demonstrating the application of the interactive prompt methodology to craft a learning scenario. This case study will showcase the step-by-step application of our structured dialogic approach, highlighting how it can be tailored to the specific demands of educational settings. 

\subsection{Implementing the Learning Scenario Assistant}
Leveraging the structured methodology outlined earlier, the interactive prompt was implemented to facilitate a rich, AI-mediated dialogue for designing educational learning scenarios. The development unfolded methodically, encompassing a series of distinct but interconnected stages:

\subsubsection{Stage 1: Defining the Position Prompt}
The initial step involved establishing the context for the LLM’s operation by utilizing a positioning prompt. This served to position the LLM within the domain of lesson plan creation, outlining best practices to guide its understanding. The prompt was crafted to be concise, limiting the response to 200 words to ensure focus and relevance. This step oriented the LLM within the educational design context, setting the stage for more targeted inquiries.

\subsubsection{Stage 2: Crafting the Interactive Prompt}
Following the LLM’s positioning process, the interactive prompt was introduced. This step involved presenting a scenario in which the GenAI tool assumes the roles of a teaching assistant, instructional designer, and subject expert simultaneously. This multi-faceted role-play approach is designed to provide a well-rounded perspective on lesson plan development, integrating various professional insights. The prompt captures the task's complexity, ensuring that the lesson plan adheres to educational standards and fosters critical thinking and comprehensive knowledge.

\subsubsection{Stage 3: Structured Follow-up Prompts}
The follow-up prompts are constructed as a series of questions, each deepening into the specific requirements of the lesson plan. These prompts facilitate a conversational exchange where the AI gathers incremental details about the target audience, Topic, goal, format, duration, and examples. The AI simulates an engaging dialogue by requesting user responses, ensuring the lesson plan is tailored to the user's needs.

\subsubsection{Stage 4: Iterative Development and Evaluation}
The methodology underscores an iterative process, where the AI presents initial outputs and actively seeks user’s feedback to refine the lesson plan. This collaborative process mimics a real-world design cycle, where initial drafts are subject to review and revision. The LLM is instructed to encourage users to revise and work on to improve a first draft of the lesson plan produced by the LLM, or request regeneration of the output, demonstrating adaptability in the AI's approach.

\subsubsection{Stage 5: Finalization and Human Touch}
The culmination of this interactive prompt sequence leads to the practical application and human-led refinement of the AI-generated lesson plan. During this phase, the AI assists users in editing and personalizing the plan, providing a final check for language and addressing major concerns. This phase emphasizes the benefits of human-AI synergy, where human creativity and expertise complement the AI's computational efficiency.

The final version of the Learning Scenario Assistant is this: 
\emph{1. Positioning Prompt} \\
“Do you know any good practices for creating a lesson plan? Please provide bullet points outlining the key practices you would use to create a sound and comprehensive lesson plan (200 words max).''

\emph{2. Interactive Prompt}\\
Role: You are a team of a teaching assistant, an experienced instructional designer, and a subject expert on the topic that will be taught.

Task: Your job is to work collaboratively to develop a comprehensive lesson plan that introduces the Target Audience to crucial concepts related to the Topic. The lesson plan should align with state educational standards to cultivate students' knowledge and critical thinking while checking for understanding.

First, introduce yourself and let the user know you will ask them questions to create a lesson plan tailored to their needs. Ask question number 1 from the list below and wait for the user to respond in a follow-up prompt. Then, move to question 2. Wait for user to respond, etc.

Here you have the questions on the following topics:
\begin{enumerate}
    \item Target Audience: Ask: "Who is your target audience (e.g., primary school students, high school students)?''
    \item Topic/Project Context: Ask the following question: “What is the subject (e.g., Biology, Physics) and the specific topic you are going to teach (e.g., ecological awareness)?''
    \item Goal: Ask the following question: "What is your ultimate goal (e.g., The ultimate goal of this lesson plan is to foster ecological literacy, encouraging students to recognize the interdependence between human actions and the environment/By the end of the lesson, students should be able to identify key ecological concepts and apply them to real-life scenarios)?''
    \item Format: Ask the following question: “What do you want the lesson plan to include (e.g., clear learning objectives, interactive activities, multimedia resources, interleaving learning, flipped classroom, assessment methods, differentiation strategies, effective learning, knowledge retention)?''
    \item Duration: Ask the following question: "How long will your lesson be?''
    \item Examples: Ask the following question: "Do you have any good examples of lesson plans you would like me to use as a template?''
    \item Additional Info: Ask the following question: "May I ask some additional questions that will help me structure your lesson plan better? (Respond with YES/NO)'' If the user responds with YES, based on the responses you have collected in questions 1 to 6, detect missing info or valuable information you would need to collect to create a better lesson plan, and ask a maximum of 2 questions that will help you clarify this.
\end{enumerate}

Follow the next steps to create the lesson plan:
\begin{enumerate}[label=\alph*]
    \item Keep in mind what you now know about the user to customize the lesson plan you will create.
    \item Start by saying: "This is the first version of a lesson plan I created based on your input. I am here to work collaboratively with you to revise it and reach the desired output.''
    \item Provide the user with a detailed lesson plan based on the information you have.
    \item Ask the user: "Are you generally happy with the first draft of the lesson plan, or should I remake it from scratch? (Respond with: CONTINUE/REGENERATE)'' If they respond CONTINUE, move on to the next step of the instructions; if they respond REGENERATE, repeat step d.
    \item Ask the user: "What would you like to improve/change/adjust in the lesson plan I created?'' Wait for the user to respond, thank them briefly, and make adjustments accordingly.
    \item Ask the user: "Is there anything else you would like me to improve? (Respond only with YES/NO)'' If they respond YES, repeat step e; if they respond NO, proceed to the next step.
\end{enumerate}

Follow the next steps to revise the human-edited lesson plan:
\begin{enumerate}[label=\alph*]
    \item Start by saying: "Now it is your time to review and edit the lesson plan created to add your personal touch and expertise. Once you are ready, please copy-paste the edited text so that I can final check for any spelling mistakes or suggest major corrections.'' Wait for the user to paste the edited lesson plan.
    \item Revise the edited lesson plan and highlight suggested changes. Refrain from making major changes. ONLY check English and make major corrections if needed.
\end{enumerate}

\subsection{Example Lesson Plan}
We tested the GenAI tool to create a lesson plan for a course on \emph{Digital Humanities} taught by K. Karpouzis, University Professor in the Department of Communication, Media, and Culture, Panteion University of Social and Political Sciences. The main aim of this experiment was to test the adaptivity of the generated results, based on the successive refinement prompts entered by the user, to illustrate its value in producing personalized learning experiences. The initial target audience definition was:
\begin{verbatim}
    150 university freshmen of a school
    of Communication, Media and Culture.
    They have a strong background
    on social sciences and humanities,
    but little hands-on experience on
    technological and computing-related
    concepts
\end{verbatim}

The prompt for the learning objectives question was:
\begin{verbatim}
    Students should be able to design
    and evaluate digital services
    related to Digital Humanities (DH),
    ask qualitative questions, then
    utilize quantitative methods, and
    then evaluate and critically discuss
    the quantitative results, discuss
    and utilize open data, and AI
    techniques and limitations
\end{verbatim}

The next interaction concerned specific elements, components or activities that the generated lesson should incorporate. We asked for
\begin{verbatim}
     1-2 team projects for the course,
     a set of self-assessment questions
     for each lecture, and additional
     video or e-book resources for each
     lecture
\end{verbatim}

Upon receiving a request for information regarding the organization's structure in terms of hours and class sessions, we were prompted to either provide exemplary lesson plans for use as templates or to design new lesson plans from scratch. For this purpose, we utilized J. Drucker's book \cite{b17}, which comprehensively covers most of the themes we aimed to incorporate into our lesson plan. Below is the first draft of the lesson plan:\ 

\textbf{\emph{Digital Humanities Lesson Plan}}

\textbf{Target Audience:} 1st-year university students in Communication, Media, and Culture with a strong background in social sciences and humanities.
\textbf{Lesson Objective:} By the end of this course, students should be able to design and evaluate digital services related to Digital Humanities (DH), ask qualitative questions, utilize quantitative methods, critically discuss the quantitative results, discuss and utilize open data, and understand AI techniques and limitations.\\
\textbf{Course Duration: }Each lesson is approximately 2 hours long.\\
\textbf{Lesson Components:}
\textbf{Team Projects (2): }
Students will work in teams to design and evaluate digital services related to Digital Humanities.\\
\textbf{Self-Assessment Questions: }
For each lecture, a set of self-assessment questions will be provided to help students gauge their understanding.\\
\textbf{Additional Resources: } Each lecture will include supplementary video or e-book resources from "The Digital Humanities Courseboo'' by Johanna Drucker to enhance learning.\\

\textbf{\emph{Lesson Structure:}}\\
\textbf{Lecture 1:} Introduction to Digital Humanities and its Scope\\
\textbf{Lecture 2:} Qualitative Research in DH\\
\textbf{Lecture 3:} Quantitative Methods in DH\\
\textbf{Lecture 4:} Data Analysis and Interpretation\\
\textbf{Lecture 5:} Open Data and Data Sharing\\
\textbf{Lecture 6:} AI in Digital Humanities and Its Limitations

This initial generation served as a preliminary step towards creating a comprehensive learning plan. Our GenAI tool is specifically designed to interact with users and fine-tune its outputs after gathering additional information. In this phase, we requested a revision considering the specified number of lectures (10 lectures per course, plus additional weeks for revision and invited talks). This revision led to a curriculum with a stronger emphasis on 'Artificial Intelligence in the Humanities' and 'Open Data.' In the second iteration, we sought to enrich the content with topics related to social media, fake news, and fact-checking. This led to the addition of lectures on: \textbf{Ethical Considerations in Digital Media}, \textbf{Media Consumption and Audience Behavior}, and \textbf{Digital Activism and Social Change}.

For the final set of suggestions, the GenAI tool suggested themes for the group activities or projects that the course should include and the necessary software and data to be processed:

\begin{itemize}
    \item \textbf{Social Media and Mental Health:} Investigate the impact of social media on mental health and well-being. Explore using specific platforms (e.g., Instagram, TikTok) and their effects on users' mental health.

    \item \textbf{Fake News and Disinformation Campaigns:} Analyze the spread of fake news and disinformation on social media platforms. Investigate the role of algorithms, user behavior, and potential solutions for fact-checking and countering disinformation.

    \item \textbf{Digital Activism and Online Movements:} Study digital activism and online social movements. Examine recent or historical examples of digital campaigns influencing social and political change.

    \item \textbf{Media Representation and Bias:} Explore the portrayal of gender, race, and other identities in digital media. Analyze how biases and stereotypes are perpetuated or challenged through digital content.

    \item \textbf{Digital Storytelling and Narrative Analysis:} Analyze digital narratives through blogs, podcasts, or YouTube channels. Investigate the power of storytelling in shaping opinions and beliefs.
\end{itemize}

with the software suggestions, including social media tools, e.g., Hootsuite and Buffer, Content Management Systems, such as WordPress, and text analysis software (Voyant Tools). Going one step further, the GenAI tool listed specific activities for text analysis (e.g., Word Frequency Analysis or Topic Modelling, both important concepts in DH research).


\section{Evaluation of Learning Scenarios}
In this section, the evaluation of the learning scenarios will be presented. 

\subsection{Evaluation procedure}

The preceding section outlined the methodology for developing the Learning Scenario Assistant. This section shifts focus to the evaluation process, designed to assess the prompt's effectiveness using a combination of quantitative and qualitative metrics. The evaluation involves a series of comprehensive steps to ensure thoroughness and reliability. The evaluation procedure involved the following steps:
\subsubsection{Testing Across Diverse Large Language Models}
The evaluation process involved testing the effectiveness of these prompts across three different Large Language Models (LLMs): ChatGPT, LLama, and Google Bard. This step was essential to gauge the adaptability and efficiency of the prompts across various AI platforms, each with its unique processing capabilities.

\subsubsection{Quantitative Analysis}
The core of the quantitative evaluation lies in scoring each LLM's response against predefined criteria. We selected these criteria to capture key elements of educational content generation, focusing on relevance and accuracy, creativity and engagement, personalization, coherence, flow, and response time. A Likert scale ranging from 1 to 5 was used for this purpose, where higher scores indicated superior performance. This structured scoring system provided an objective framework for measuring and comparing the capabilities of each LLM in responding to the educational prompts.

\subsubsection{Qualitative Feedback}
Complementing the quantitative analysis was the qualitative evaluation, which involved gathering insights from teachers and educational experts. They were asked to provide detailed feedback and additional scores for each AI-generated response. This step was crucial in capturing the more nuanced aspects of the content, such as its pedagogical value, engagement level, and overall suitability for educational purposes.

\subsubsection{Linguistic Analysis}
In the Linguistic Analysis stage, we focused on examining the language used by each Large Language Model (LLM). Using spaCy, we analyzed the structure of the text, breaking it down into tokens to understand the models' use of vocabulary and grammar. Additionally, we applied Latent Dirichlet Allocation (LDA) to identify the main themes in the AI-generated responses. This step was crucial in evaluating the clarity, relevance, and thematic accuracy of the content provided by each LLM, ensuring that their responses were correct, contextually appropriate, and engaging for educational purposes.

\subsubsection{Comprehensive Evaluation Outcome}
The amalgamation of quantitative and qualitative assessments yielded a holistic view of the performance of each LLM. This comprehensive approach was instrumental in achieving a balanced evaluation, effectively capturing both the tangible, measurable aspects of the AI responses and the subtler, qualitative elements vital in education.

The Learning Scenario Assistant underwent testing in preschool, secondary school, and university levels, as well as in English and Greek. It generated tailored responses for each educational level, allowing an evaluation of its adaptability and versatility in meeting diverse academic requirements.

The evaluation encompassed two key aspects: a quantitative analysis and a qualitative assessment. The quantitative analysis was conducted by experts who scored each language model's (LLM) response against predefined criteria. Concurrently, the qualitative assessment involved gathering feedback from educational professionals to gauge the practical applicability and relevance of the responses.

A critical component of our methodology was the integration of a Linguistic Analysis stage. We utilized spaCy for detailed tokenization and employed Latent Dirichlet Allocation (LDA) to analyze the textual data. LDA was instrumental in uncovering the predominant themes within the LLM responses, thereby assessing their thematic relevance and accuracy within the context of the specified educational scenarios.

The combination of these varied yet interrelated evaluation methods culminated in a comprehensive analysis. This approach not only quantified the performance of each LLM but also qualitatively appraised the more subtle, pedagogically significant elements of the generated content. By adopting this holistic evaluation strategy, we ensured a balanced and thorough assessment, capturing both the measurable performance and the qualitative nuances of the AI-generated responses, which are crucial for their application in educational settings.

\subsection{Evaluation Results}
The evaluation results are presented in the following tables.
\begin{table}
    \centering
    \begin{tabular}{p{1.4cm}p{0.7cm}p{0.7cm}p{0.7cm}p{0.7cm}p{0.7cm}p{0.8cm}}
         &  \multicolumn{6}{c}{GenAI models
}\\
         &  \rot{ChatGPT 3.5 (09/2023)}&  \rot{ChatGPT 4 (09/2023)}&  \rot{Llama 2 7B (07/2023)}&  \rot{Llama 2 13B (07/2023}&  \rot{Llama 2 70B (07/2023)}& \rot{Google Bard (2023.09.27)}\\\hline
         Relevance&  4.66&  5&  4&  4.66&  5& 4
\\\hline
         Accuracy&  3.66&  4.66&  4&  4&  3& 4
\\\hline
         Creativity&  3.33&  4.66&  4.33&  5&  5& 3.33
\\\hline
         Engagement&  3.33&  5&  4&  4.33&  5& 3
\\\hline
 Personalization& 3.66& 5& 3.66& 4.33& 5&3
\\\hline
 Coherence& 5& 5& 4.66& 5& 5&3
\\\hline
 Response Time& 4.66& 3.66& 4.66& 5& 3.33&5
\\\hline
    \end{tabular}
    \caption{Quantitative Evaluation for English Learning Scenarios }
    \label{tab:quant_eval}
\end{table}
Regarding the quantitative evaluation metrics, ChatGPT 4 (September 2023) and Llama 2 70B (July 2023)' generally scored the highest across most categories, indicating their superiority in terms of relevance, accuracy, creativity, and personalization. However, Llama 2 13B (July 2023) excels in creativity and response time. It's noteworthy that Google Bard, while competitive, tends to score lower in several categories like coherence and engagement. Regarding accuracy, Llama 2 70B (July 2023) scored relatively poorly as it provided material that was not existent (e.g., links for additional resources that did not exist).

\begin{table}
    \centering
    \begin{tabular}{p{1.7cm}p{0.7cm}p{0.7cm}p{0.7cm}p{0.7cm}p{0.7cm}p{0.8cm}}
         &  \multicolumn{6}{c}{GenAI models
}\\
         &  \rot{ChatGPT 3.5 (09/2023)}&  \rot{ChatGPT 4 (09/2023)}&  \rot{Llama 2 7B (07/2023)}&  \rot{Llama 2 13B (07/2023}&  \rot{Llama 2 70B (07/2023)}& \rot{Google Bard (2023.09.27)}\\\hline
         Value Relevance&  4.33&  4.66&  4&  4&  4.33& 3
\\\hline
         Understandability&  4.33&  5&  3.66&  4&  5& 3
\\\hline
         Measurability&  4.33&  4.66&  3.66&  3.66&  4.33& 4
\\\hline
         Non-redundancy&  4.66&  4.66&  4.33&  4.66&  4.66& 3.66
\\\hline
 Judgmental Independence& 5& 5& 5& 5& 5&4
\\\hline
 Balancing Completeness and Conciseness& 4.33& 4.33& 4& 3.66& 4.66&4.33
\\\hline
 Operationality& 5& 5& 4.33& 4& 4.33&3.66
\\\hline
 Simplicity vs. Complexity& 4& 5& 3.66& 4& 4.66&4.33\\
    \end{tabular}
    \caption{Qualitative Evaluation for English Learning Scenarios}
    \label{tab:qual_eval}
\end{table}
Regarding the qualitative evaluation metrics, ChatGPT 4 (September 2023) consistently scores high across all categories, suggesting it provides relevant, understandable, measurable, and operationally practical outputs. Llama 2 70B (July 2023) also performs well in several categories, particularly in balancing completeness, conciseness, and simplicity vs. complexity. While competitive, Google Bard (2023.09.27) tends to score lower in several categories, indicating areas where it may not perform as strongly as the other models.
\begin{table}
    \centering
    \begin{tabular}{p{1.4cm}p{0.7cm}p{0.7cm}p{0.7cm}p{0.7cm}p{0.7cm}p{0.8cm}}
         &  \multicolumn{6}{c}{GenAI models
}\\
         &  \rot{ChatGPT 3.5 (09/2023)}&  \rot{ChatGPT 4 (09/2023)}&  \rot{Llama 2 7B (07/2023)}&  \rot{Llama 2 13B (07/2023}&  \rot{Llama 2 70B (07/2023)}& \rot{Google Bard (2023.09.27)}\\\hline
         Tokens&  300.33&  421&  322&  354&  484& 337.66
\\\hline
         Number of Main Topics based on LDA&  9.33&  10&  9.33&  10&  10& 8.33
\\\hline
    \end{tabular}
    \caption{Linguistic Analysis for English Learning Scenarios}
    \label{tab:ling_analysis}
\end{table}
The differences in the number of tokens might reflect varying styles and content depth among the models. Models generating more tokens might provide more thorough or detailed answers. The number of main topics based on LDA suggests how diverse or varied the responses are regarding the subject matter. Models identifying more main topics could better address a wide range of topics within a given scenario. ChatGPT 4 and Llama 2 70B stand out for generating more tokens and covering a broader range of topics, which could indicate their advanced capabilities in developing detailed and diverse content.

\subsubsection{Evaluation Results for Greek Learning Scenarios}
\begin{table}
    \centering
    \begin{tabular}{p{1.4cm}p{0.7cm}p{0.7cm}p{0.7cm}p{0.7cm}p{0.7cm}p{0.8cm}}
         &  \multicolumn{6}{c}{GenAI models
}\\
         &  \rot{ChatGPT 3.5 (09/2023)}&  \rot{ChatGPT 4 (09/2023)}&  \rot{Llama 2 7B (07/2023)}&  \rot{Llama 2 13B (07/2023}&  \rot{Llama 2 70B (07/2023)}& \rot{Google Bard (2023.09.27)}\\
         Relevance&  4.33&  5&  4&  4.66&  5& 4
\\\hline
         Accuracy&  3.33&  4.33&  3&  4&  3& 3.66
\\\hline
 Creativity& 3& 4& 3.33& 4.33& 4.66&3
\\\hline
 Engagement& 3.33& 4.33& 3.33& 4& 4.66&3
\\\hline
 Personalization& 3.66& 4.33& 3& 4.33& 4.66&3
\\\hline
 Coherence& 4/66& 4.66& 4& 4& 4.66&3
\\\hline
 Response Time& 4& 3.33& 3.33& 4& 3.33&4.66
\\\hline
    \end{tabular}
    \caption{Quantitative Evaluation for Greek Learning Scenarios}
    \label{tab:quant_eval_greek}
\end{table}
The most advanced versions of ChatGPT and Llama 2 (ChatGPT 4 and Llama 2 70B), generally score higher across most categories, suggesting they are better adapted to handle Greek language scenarios in terms of relevance, creativity, engagement, and personalization. Google Bard shows a notable strength in response time but scores lower in coherence, creativity, and personalization, indicating a potential area for improvement in handling Greek scenarios.

In some aspects, like coherence and engagement, the performance of the newer models in Greek is on par or slightly lower than in English, which might reflect the challenges associated with processing and generating content in a language different from English. Overall, these results suggest that while there is a high level of capability across these models in handling Greek language scenarios, there are variances in their effectiveness, particularly in creativity, engagement, and coherence.

\begin{table}
    \centering
    \begin{tabular}{p{1.4cm}p{0.7cm}p{0.7cm}p{0.7cm}p{0.7cm}p{0.7cm}p{0.8cm}}
         &  \multicolumn{6}{c}{GenAI models
}\\
         &  \rot{ChatGPT 3.5 (09/2023)}&  \rot{ChatGPT 4 (09/2023)}&  \rot{Llama 2 7B (07/2023)}&  \rot{Llama 2 13B (07/2023}&  \rot{Llama 2 70B (07/2023)}& \rot{Google Bard (2023.09.27)}\\
         Judgmental Independence&  3.66&  4.33&  4.33&  4&  4.33& 3.66
\\\hline
         Balancing Completeness and Conciseness&  3.66&  4.33&  3.33&  3.66&  4& 4
\\\hline
 Operationality& 4.33& 4.66& 4.33& 3.66& 4&3.66
\\\hline
 Simplicity vs. Complexity& 3.66& 4.33& 3.33& 3.66& 4.66&4.33
\\\hline
    \end{tabular}
    \caption{Qualitative Evaluation for Greek Learning Scenarios}
    \label{tab:qual_eval_greek}
\end{table}

ChatGPT 4 (September 2023) consistently scores high across most categories, suggesting its effectiveness in delivering relevant, understandable, measurable, and operationally practical outputs in Greek. Llama 2 70B (July 2023) also performs well, particularly in balancing simplicity with complexity. While competitive, Google Bard (2023.09.27) tends to score lower in several categories, indicating areas where it may not perform as strongly in Greek scenarios. These results highlight the capabilities of the more advanced models, like ChatGPT 4 and Llama 2 70B, in handling the specificities of the Greek language in learning scenarios. The variations in scores across different models suggest differing levels of adaptability and proficiency in non-English contexts.
\begin{table}
    \centering
    \begin{tabular}{p{1.4cm}p{0.7cm}p{0.7cm}p{0.7cm}p{0.7cm}p{0.7cm}p{0.8cm}}
         &  \multicolumn{6}{c}{GenAI models
}\\
         &  \rot{ChatGPT 3.5 (09/2023)}&  \rot{ChatGPT 4 (09/2023)}&  \rot{Llama 2 7B (07/2023)}&  \rot{Llama 2 13B (07/2023}&  \rot{Llama 2 70B (07/2023)}& \rot{Google Bard (2023.09.27)}\\\hline
         Tokens&  267&  392.66&  294.66&  322&  430.33& 308.33
\\
\hline
         Number of Main Topics based on LDA&  9.33&  10&  9&  10&  10& 8.66
\\
\hline
    \end{tabular}
    \caption{Linguistic Analysis for Greek Learning Scenarios }
    \label{tab:ling_analysis_greek}
\end{table}
In English and Greek scenarios, newer models like ChatGPT 4 and Llama 2 70 generally outperform others regarding detailed responses and topic diversity. There's a notable consistency in the performance of models across English and Greek, suggesting a level of adaptability in different linguistic contexts. However, the slightly reduced token count and topic range in Greek indicate challenges or differences in handling non-English languages. In addition, there's a clear trend where more advanced models provide more comprehensive (in terms of length) and diverse (in terms of topics) responses. However, the extent of this comprehensiveness and diversity appears slightly moderated in Greek.

In summary, while the leading models demonstrate strong capabilities in both English and Greek, there are subtle differences. These might stem from the intrinsic challenges of processing and generating content in a less commonly modeled language like Greek, compared to English, which has traditionally been the primary focus of language model training and optimization.

\section{Discussion}
The advent of Generative AI (GenAI) in education, especially its ability to create individualized lesson plans, represents a significant shift from traditional teaching methods. GenAI's tailored approach, addressing each student's unique learning style and pace, offers a more effective and inclusive learning environment. Our study employed a comprehensive evaluation methodology, integrating both quantitative and qualitative measures, to assess GenAI's effectiveness across various parameters. These included relevance, accuracy, creativity, student engagement, and response time, providing a holistic view of its impact in educational settings.

Preliminary results indicate that educators find GenAI-generated lesson plans not only effective but also a significant time-saver in lesson preparation. These customized plans cater to diverse learning needs, enhancing the educational experience and fostering adaptability. The introduction of an interactive mega-prompt methodology in lesson planning represents further innovation, offering a flexible foundation adaptable to various learning styles and preferences. However, this innovative approach comes with challenges, particularly in ensuring AI's quality, ethical use, and accuracy in educational contexts. These concerns underscore the need for continuous oversight and evaluation of AI tools in education. Ongoing oversight of AI tools in education is essential. Collaboration between educational entities and regulatory authorities is key for formulating policies that support ethical AI use in education, following UNESCO's guidelines for generative AI \cite{b16}.

Recognizing the necessity for broader, more extensive testing to explore GenAI's scalability and efficacy in diverse educational contexts, including special education needs, Randomized Control Trials (RCTs) emerge as a pivotal tool. RCTs provide an unbiased, comparative framework to evaluate various GenAI tools' effectiveness in different learning environments. Drawing from critical methodological developments discussed in the work of M. Pampaka et al. \cite{m9}, RCTs can address some of the complexities and nuances highlighted in the paper. As we move into this new horizon of AI-driven education, balancing AI assistance with human oversight becomes imperative. It is essential to continually assess the impact of AI on educational systems and consider the broader implications for global education standards.

\section{Conclusion}
This paper has demonstrated the revolutionary role of Generative AI (GenAI) in transforming educational content generation. We have established its adaptability across various educational levels and languages, highlighting its ability to enhance learning through personalized and engaging materials. Our comprehensive analysis, encompassing both quantitative and qualitative assessments, indicates the effectiveness of GenAI in education. This research underscores AI's potential to enhance educational practices and paves the way for future developments in learner-centric educational approaches and technologies. As the field of AI in education continues to evolve, ongoing research and exploration are essential. This will enable further refinement of AI tools, amplifying their benefits and ensuring their effective integration into diverse educational settings.

\section*{Acknowledgment}
This work has been funded by the Horizon Europe Road-STEAMER project (Grant number: 101058405). More information can be found at https://www.road-steamer.eu  

\balance

\vspace{12pt}

\end{document}